\documentclass[aps,prd,groupedaddress,amsmath,amssymb,showpacs,
reprint
]{revtex4-1}

\usepackage{graphicx}    
\usepackage{multirow}    
\usepackage{rotating}    
\usepackage{dcolumn}     
\usepackage{bm}          
\usepackage{hyperref}    

\newcommand{\beq}{\begin{equation}}
\newcommand{\eeq}{\end{equation}}
\newcommand{\bea}{\begin{eqnarray}}
\newcommand{\eea}{\end{eqnarray}}
\newcommand{\rmd}{{\mathrm d}}
\newcommand{\mrm}[1]{\mathrm{#1}}

\newcommand{\eref}[1]{(\ref{#1})}

\newcommand{\ber}{\begin{sideways}}
\newcommand{\eer}{\end{sideways}}
\newcommand{\dgr}{\dagger}
\newcommand{\bigO}{\mathcal{O}}
\newcommand{\Gbar}{\overline{G}}
\newcommand{\Qbar}{\overline{Q}}
\newcommand{\Pbar}{\overline{P}}
\newcommand{\Sbar}{\overline{S}}

\newcommand{\Tr}{\mathrm{Tr}\;}
\renewcommand{\Re}{\mathrm{Re}\;}
\newcommand{\abs}[1]{\lvert#1\rvert}
\newcommand{\norm}[1]{\lVert#1\rVert}

\newcolumntype{d}[1]{D{.}{.}{#1}}

\begin{document}


\title{General heatbath algorithm for pure lattice gauge theory}


\author{Robert W. Johnson}
\email[]{robjohnson@alphawaveresearch.com}
\homepage[]{http://www.alphawaveresearch.com}
\affiliation{Alphawave Research, Atlanta, GA, 30238, USA}


\date{\today}

\begin{abstract}
A heatbath algorithm is proposed for pure SU($N$) lattice gauge theory based on the Manton action of the plaquette element for general gauge group $N$.  Comparison is made to the Metropolis thermalization algorithm using both the Wilson and Manton actions.  The heatbath algorithm is found to outperform the Metropolis algorithm in both execution speed and decorrelation rate.  Results, mostly in $D=3$, for $N=2$ through 5 at several values for the inverse coupling are presented.
\end{abstract}


\maketitle


\section{\label{sec:intro}Introduction}
A heatbath algorithm is proposed for pure SU($N$) lattice gauge theory based on the Manton action of the plaquette element~\cite{Manton:1980328}.  The relation of the Wilson and Manton actions is equivalent to the relation of the Frobenius and Riemann metrics for the distance between elements of the gauge group~\cite{moakher:02895}.  While heatbath efficiency has been achieved for the cases of SU(2) and U(1) using a biased-Metropolis algorithm~\cite{bazavov:114506,bazavov:117501}, and a heatbath algorithm for SU(2) is known~\cite{Creutz:1980mc,Kennedy:1985nu} which may be extended to $N>2$ using covering subgroups~\cite{Cabibbo1982387}, the composition of a direct heatbath updating scheme for general gauge group $N$ has been an outstanding problem in lattice gauge theory for quite some time.  As the direct approach seems intractable, we consider an indirect means to accomplish the thermal updating.  By relating the Riemann norm of a plaquette in the matrix representation $Q$ to the Frobenius norm of the plaquette in the vector representation $q$, one may generate random plaquette elements of known Manton action, which reduces to the Yang-Mills action in the continuum limit.  The algorithm here generates a proposal for the updated mean plaquette by averaging some number of random elements with action drawn from a gamma distribution.  Dividing out the mean environment contribution then provides the updated link value.

After calibration against known results~\cite{Teper:1999}, we compare the performance of the proposed heatbath algorithm to the Metropolis thermalization algorithm~\cite{metropolis:1087} using both the Wilson and Manton actions~\cite{Montvay:1994cy}.  We show that one may generate identical lattice action distributions using either thermal updating scheme, with demonstration for a $D=3$ and $L=4$ isotropic lattice having periodic boundary conditions.  The running time of the heatbath algorithm compares favorably with that of the Metropolis algorithm using either action and displays the expected linear dependence on lattice volume $L^D$.  Both the link and plaquette decorrelation rates are evaluated for a particular case of $N$ and $\beta$, where the heatbath algorithm is found to outperform the Metropolis algorithm in both execution speed and decorrelation rate.

Summary of notation: \\
Isotropic Euclidean lattice regularization $x/a \in \{L^D\}$\\
Special unitary group SU($N$) with identity $\Tr I = N$\\
Link variables $U \in \{U\}$, plaquette variables $Q \in \{Q\}$\\
Exponential parametrization $Q = \exp (i \sum_a q_a H_a)$\\
Generator normalization $\Tr H_a H_b = 2 \delta_{ab}$\\
Conjugate transposition $U^\dgr = U^{-1}$, $H^\dgr = H$\\
Frobenius norm $\norm{Q}_F^2 = \Tr Q^\dgr Q$, $\norm{q}_F^2 = \sum_a q_a^2$\\
Frobenius metric $\rmd_F(Q_1,Q_2) = \norm{Q_1 - Q_2}_F$\\
Riemann norm $\norm{Q}_R^2 = \norm{\log Q}_F^2 / 2$\\
Riemann metric $\rmd_R(Q_1,Q_2) = \norm{\log (Q_1^\dgr Q_2)}_F / \sqrt{2}$\\
Boltzmann factor $\exp(-S) = \exp(- \beta \sum_Q S_Q)$\\
Wilson action $S_W(Q) = \rmd_F^2(I,Q) / 2 N = 1 - \Re \Tr Q / N$\\
Manton action $S_M(Q) = \rmd_R^2(I,Q) / N = \norm{\log Q}_F^2 / 2 N $

\section{\label{sec:action}Comparison of action definitions}
The Wilson action is recognized as a measure of the gauge invariant plaquette element's distance from the identity normalized by the gauge group $N$.  Working backwards from the usual expression, one writes \bea
S_W(Q) &=& 1 - \Re \Tr Q / N \;, \\
 &=& \Tr \left[ 2 I - \left( Q + Q^\dgr \right) \right] / 2 N \;, \\
 &=& \Tr \left[ \left( I - Q \right)^\dgr \left( I - Q \right)\right] / 2 N \;, \\
 &=& \rmd_F^2(I,Q) / 2 N  \;,
\eea where the Frobenius metric measures the distance from the identity to the plaquette in the space of the general linear group GL($N,\mathbb{C}$).  Similarly, the Manton action is related to the Riemann metric of the plaquette matrix $Q$, which measures the distance along the one parameter subgroup of SU($N$) connecting $I$ to $Q$, given by $Q^\rho \equiv \exp (\rho \log Q)$ for $0 \leq \rho \leq 1$.  Using the exponential parametrization $Q = \exp(i \sum_a q_a H_a)$ for traceless Hermitian generators $H$ normalized to $\Tr H_a H_b = 2 \delta_{ab}$, the Riemann norm in the matrix representation is equal to the Frobenius norm in the vector representation, $\norm{\log Q}_F^2 = \Tr \sum_a q_a^2 H_a^2 = 2 \norm{q}_F^2$, so that the Manton action is proportional to squared radius of the parameter vector $q \equiv r \hat{q}$, giving $S_M(Q) \propto \norm{Q}_R^2 = r^2$.  The maximum for either action is given by its value for an element antipodal to the identity $S_Q(I')$, where $I' \equiv \mrm{diag}[-1, \ldots, -1, \pm1]$ and the final sign ensures $\mrm{det} I' = 1$, yielding $S_W(I') = 4 \lfloor (N/2) / N$ and $S_M(I') = \lfloor (N/2) \pi^2 / N$ for $\lfloor (N/2)$ the greatest integer $\leq N/2$.

\begin{figure}
\includegraphics[]{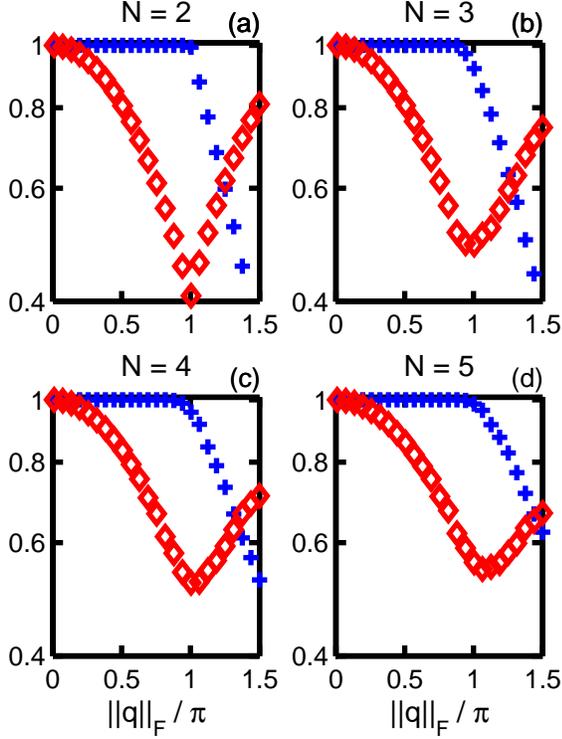}%
\caption{\label{fig:A} Displayed as $+$ is the ratio of the matrix and vector norms $\norm{Q}_R / \norm{q}_F$ averaged over 100 random unit vectors $\hat{q}$ for various $N$.  Also shown as $\Diamond$ is the ratio of the Wilson and Manton actions $S_W(Q) / S_M(Q)$ which approaches unity as $\norm{q}_F \rightarrow 0$.  The injective limit $\norm{q}_F = \pi$ corresponds to an element antipodal to the identity.}
\end{figure}

The exponential parametrization of SU($N$) elements is unique out to the injective limit $r < \pi$---beyond that, the matrix $i \sum_a q_a H_a$ is not the principle logarithm of the corresponding plaquette $Q$, indicating that a shorter path through the group connects the element to the identity, as seen in Fig.~\ref{fig:A}.  For reasonable values of $\beta$ close to the continuum limit, the chance of encountering elements near the injective limit is exponentially suppressed.  Writing the logarithm as $\log Q = - \sum_{k=1}^\infty (I - Q)^k / k$, we see that the Wilson action corresponds to taking only the first term in the expansion.  Consequently, it may be viewed as the leading order approximation to the Manton action, similar to the relationship between the chordal and arc distance between two points on a circle.  In the limit of vanishing action $S_Q \rightarrow 0$, the two expressions become equal, so that the ratio $S_W(Q) / S_M(Q) \rightarrow 1$, also shown in Fig.~\ref{fig:A}---at finite $\beta$, one expects $S_W(Q) < S_M(Q)$.

The equivalence of the action definitions in the continuum limit may be shown analytically.  The relation of the Wilson action to the Yang-Mills action is well known~\cite{Montvay:1994cy} and will not be repeated here.  Picking up the derivation for the Manton action in $D = 4$ at the plaquette discretization $Q_{\mu \nu}(x) = \exp [- a^2 G_{\mu \nu}(x)]$ for $G_{\mu \nu}(x) = F_{\mu \nu}(x) + \bigO(a)$, one finds $\log Q = - a^2 F_{\mu \nu} + \bigO(a^3)$.  Then one may write \beq
S_M(Q) = \dfrac{1}{N} \rmd_R^2(I,Q) = \dfrac{1}{2 N} \norm{\log Q}_F^2 = \dfrac{a^4}{2 N} \Tr G^\dgr_{\mu \nu} G_{\mu \nu} \;,
\eeq and doubling the plaquette sum $2 \sum_Q = \sum_{x,\mu,\nu}$ one obtains \beq
S = \beta \sum_Q S_M(Q) = - \dfrac{\beta}{4 N} \sum_x a^4 F_{\mu \nu}(x) F^{\mu \nu}(x) + \bigO(a^5) \;,
\eeq where the minus sign appears because $A_\mu(x) = - i g A_\mu^b H_b$ for coupling constant $g$.  Thus, pure lattice gauge field theory may be described equally in the limit $\beta \rightarrow \infty$ by either the Wilson or the Manton action.  The coupling may become dimensionful $\beta = 2 N / a^{4-D} g^2$ when $D \neq 4$.  The numerator for $\beta$ is conventional and serves to cancel the normalization factor in the denominator of the definition of the action; it might be more physical to normalize by the number of degrees of freedom in the gauge field $d \equiv N^2-1$ so that the normalized Manton action becomes the mean squared value of the parameters in the vector representation.

Later we will make use of the average of some number of SU($N$) elements, so let us now look at how the projection of the sum relates to the choice of action.  The link to be updated $U$ is surrounded by $n_G \equiv 2^{D-1}$ staples $G_k \in \mrm{SU}(N)$ composed of the remainder of the plaquettes involving $U$ whose sum belongs to the general linear group $\sum_k G_k = G \in \mrm{GL}(N,\mathbb{C})$.  Generalizing the discussion by Moakher~\cite{moakher:02895}, the Frobenius mean is the group element $\Gbar_F \in \mrm{SU}(N)$ which minimizes the metric of GL($N,\mathbb{C}$), such that $\Gbar_F = \arg \min \sum_k \rmd_F^2 (U, G_k)$, which one may show is equal to the projection of $G$ onto SU($N$) that maximizes $\Re \Tr \Gbar_F^\dgr G$.  The most expedient evaluation~\cite{morning:04695,Liang:1992cz} is first to project from GL($N,\mathbb{C}$) to SL($N,\mathbb{C}$) by dividing out the $N$th root of the determinant $\widetilde{G} = G / \abs{G}^{1/N}$ and then from SL($N,\mathbb{C}$) to SU($N$) using $\Gbar_F = \widetilde{G} ( \widetilde{G}^\dgr \widetilde{G} )^{-1/2}$.  Alternately, one may write $\Gbar_F = G (G^\dgr G)^{-1/2} \abs{G^{-1} G^\dgr}^{1/2N}$, requiring care that the determinant comes out 1 when $G$ is far from SU($N$).  The Riemann mean is the element $\Gbar_R$ which minimizes the metric of SU($N$), such that $\Gbar_R = \arg \min \sum_k \rmd_R^2 (U, G_k)$, which coincides with $\Gbar_F$ for the case $n_G = 2$; otherwise, the means may differ but only by some small amount for elements sufficiently close together.  No simple formula exists for $\Gbar_R$ in the general case (to our knowledge), though it may be found from a numerical minimization of the metric, and so for the following we will write $\Gbar$ for the projection of the staple sum, using $\Gbar_F$ for its evaluation when we really ought to be using $\Gbar_R$.

\begin{figure*}
\includegraphics[]{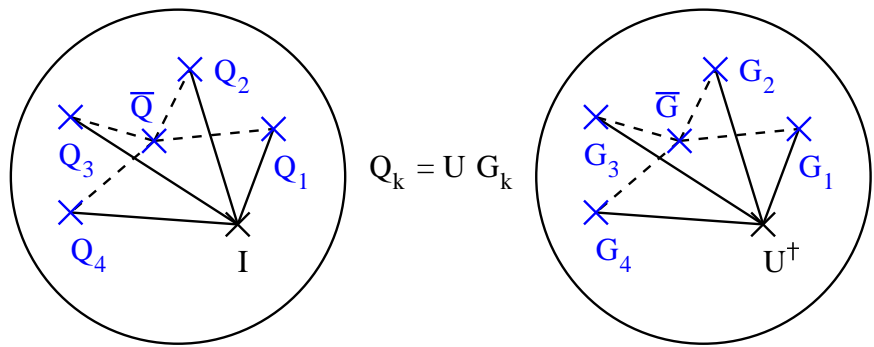}%
\caption{\label{fig:QUG} Distances are preserved under a unitary transformation of the group elements.  The environment projection $\Gbar$ minimizes the sum of the distances to the elements $G_k$, as shown by the dashed lines.  The distances appearing in the action are shown by the solid lines.}
\end{figure*}

The bi-invariance of either metric ensures that distances are preserved under a unitary transformation $\rmd (UAV,UBV) = \rmd (A,B)$, which for the Wilson action accounts for the permutation of the order of the elements under the trace.  Thus, one can relate graphically the action of the plaquettes $Q_k = U G_k$ to the distances from $U^\dgr$ to the environment elements $G_k$, as shown in Fig.~\ref{fig:QUG}.  The projection of the plaquettes is given simply by $\Qbar = U \Gbar$.  One can see that the action of the projected plaquette sum need not be simply related to the sum of the plaquette actions.

\section{\label{sec:thermal}Description of thermalization algorithms}
We will start by discussing the commonly used algorithm by Metropolis as well as the Creutz-Kennedy-Pendleton heatbath algorithm for SU(2) and why its direct extension appears intractable for $N>2$.  We will then describe our prescription for a general heatbath algorithm based on the Manton action.  We will show in the following section that identical action distributions are generated by the Metropolis and heatbath algorithms, with some allowance for the case of  $N=2$.  On a technical level, we have found that, even for the case of SU(2), while the overrelaxation algorithm~\cite{Creutz:1987xi,Adler:37458,forcrand:28504} leaves the Wilson action invariant for $U \rightarrow \Gbar_F^\dgr U^\dgr \Gbar_F^\dgr$, the Manton action is not invariant (though is nearly so for $\Gbar_R^\dgr$), and so we will not be including microcanonical overrelaxation in this study.

\subsection{\label{sec:metro}Metropolis algorithm}
The benchmark algorithm for the thermal updating of gauge field configurations according to the Boltzmann factor $\exp (-S) = \exp (- \beta \sum_Q S_Q) \equiv \prod_Q W_Q$ is that of Metropolis~\cite{metropolis:1087}, where the local transition probability for non-uniform sampling is composed of a trial and acceptance probability $p (U_\mrm{new} \leftarrow U_\mrm{old}) = p_A (U_\mrm{new} \leftarrow U_\mrm{old}) p_T (U_\mrm{new} \leftarrow U_\mrm{old})$ such that \beq \label{eqn:metropA}
p_A (U_\mrm{new} \leftarrow U_\mrm{old}) \propto \min \left[ 1, \dfrac{W_Q^\mrm{new} / p_T (U_\mrm{new} \leftarrow U_\mrm{old})}{W_Q^\mrm{old} / p_T (U_\mrm{old} \leftarrow U_\mrm{new})} \right] \;.
\eeq  The usual implementation of the multi-hit Metropolis (MP) algorithm updates a link $U_\mrm{new} \leftarrow U_\mrm{old}$ by accumulating proposals $P \in \mrm{SU}(N)$ on the left of the original link $U_\mrm{new} = P U_\mrm{old}$.  If the proposals $P$ are chosen such that inverses are equally likely $p_T (P) = p_T (P^\dgr)$, then the trial probabilities cancel in Eq.~\eref{eqn:metropA}, leaving an acceptance probability proportional to the Boltzmann ratio $p_A \propto \exp [- \beta (S_\mrm{new} - S_\mrm{old})] = \exp (- \beta \Delta_S)$, which is compared to a uniform deviate $\rho \in [0, 1]$.  Here, the $h_T = 10$ proposals $P = \exp(i \sum_a p_a H_a)$ are generated by a Gaussian deviate on the parameters $p$ in $d$ dimensions with zero mean and variance controlled by a parameter $\epsilon_p \ll 1$, which adjusts during thermalization to tune the hit ratio $h_A/h_T \approx 1/2$ then is held fixed during measurements, so that $P$ and $P^\dgr$ have equal likelihood peaked at the identity.

Using the Wilson action, one can write $\Delta_S^W = \Re \Tr (I - P) U_\mrm{old} G / N$, where $G$ is the sum of the staple elements $G_k$ surrounding the link to be updated.  For the Manton action, each term in $\sum_k S_M (Q_k)$ must be evaluated; the most efficient means is to notice that $S_M (Q) = - \sum_j (\log \lambda_j)^2 / 2 N$ where $\lambda_j$ are the eigenvalues of $Q = V_Q \Lambda_Q V_Q^\dgr$.  The additional calculational load approximately doubles the running time of the MP algorithm using the Manton action in $D=3$.  We verify our construction of the MP algorithm by comparing its results to those presented by Teper~\cite{Teper:1999} making use of the algorithm by Cabibbo and Marinari~\cite{Cabibbo1982387}.

\subsection{\label{sec:ckpalg}Creutz-Kennedy-Pendleton algorithm}
For the case of $N=2$ a direct heatbath algorithm is known~\cite{Creutz:1980mc,Kennedy:1985nu}.  The CKP algorithm makes use of two features peculiar to SU(2): the simplicity of the spectral parametrization~\cite{weigert:8739W} of group elements, $U = u_0 I + i\, \vec{u} \cdot \vec{H}$ for $\vec{H} = \vec{\sigma}$ the Pauli matrices and $u_0 = \pm (1-\norm{\vec{u}}_F^2)^{1/2}$, and the proportionality of a sum of elements to another group element, $\Gbar = G / \abs{G}^{1/2}$.  The Haar measure is reduced to $dU = du_0 (1-u_0^2)^{1/2} d^2 \Omega_u$, where $\Omega_u$ is the solid angle along $\vec{u}$, and the Boltzmann weight according to the Wilson action is $\prod_Q W_Q \propto \exp (\beta \Tr U G / 2)$.  The action is parametrized as $U G = \abs{G}^{1/2} U \Gbar = \abs{G}^{1/2} \Qbar$ for $\Qbar = q_0 I + i\, \vec{q} \cdot \vec{\sigma}$, yielding a final distribution $\int d\Qbar\, \prod_Q W_Q \propto \int_{-1}^1 dq_0 (1-q_0^2)^{1/2} \exp (\beta \abs{G}^{1/2} q_0) \int d^2 \Omega_q$.  The angular distribution may be generated from normalized Gaussian deviates as above, leaving the single parameter $q_0$ to be determined.  The Creutz algorithm generates the exponential factor from the cumulative distribution and corrects for the Haar measure using a rejection step, while the Kennedy-Pendleton variant includes a piece of the Haar measure with the exponential distribution under a suitable change of variable and corrects for the remainder with a rejection step.

\begin{figure}
\includegraphics[]{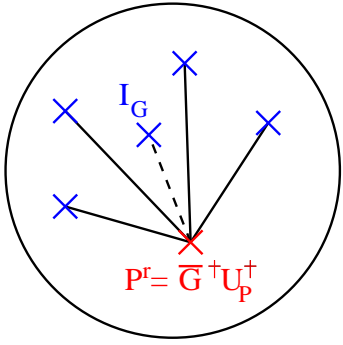}%
\caption{\label{fig:KPU} Generalization of the CKP algorithm in terms of the group metric.  The proposal $P^r$ is parametrized by its distance from the element which minimizes the action.  These elements are related to those of Fig.~\ref{fig:QUG} by $G_k = \Gbar\, (\Gbar^\dgr G_k)$.}
\end{figure}

The generalization of the CKP algorithm is depicted in Fig.~\ref{fig:KPU}, related to the right hand side of Fig.~\ref{fig:QUG} by $G_k = \Gbar\, (\Gbar^\dgr G_k)$.  The proposal is parametrized as $P^r = \exp (i r \hat{p} \cdot \vec{H})$ such that $P^0 = I$ and $P^{-r} = P^{r \dgr}$, where $\hat{p}$ is a random unit vector and $r$ is the Riemann metric between that element which minimizes the action $I_G$ and the proposal $P^r$.  Decomposing the angular dependence $P^1 = V_1 \Lambda_1 V_1^\dgr$ lets one write the sum of the actions as $S_\mrm{sum} = \sum_k S_Q (V_1 \Lambda_1^{-r} V_1^\dgr \Gbar^\dgr G_k)$, which one relates to an action value drawn from the desired distribution by $S_\mrm{draw} = S_\mrm{sum} / n_G$.  Using the Wilson action, one can write $S_\mrm{sum}^W = n_G - \Re \Tr P^{-r} (G^\dgr G)^{1/2} / N \abs{G^{-1} G^\dgr}^{1/2N}$ where $(G^\dgr G)^{1/2}$ is \textit{not} an element of SU($N$).  For the Manton action, the evaluation of the sum as a function of $r$ is even less tractable, so that no clear means of generating its distribution presents itself.  We remark, however, that a biased-Metropolis algorithm~\cite{bazavov:114506,bazavov:117501} may be built along these lines, where the distance $r$ would appear in the trial probability $p_T (P^r)$ and the distances for $S_\mrm{sum}$ appear in the Boltzmann ratio.

\subsection{\label{sec:heatbath}Indirect heatbath algorithm for general $N$}
\begin{figure}
\includegraphics[]{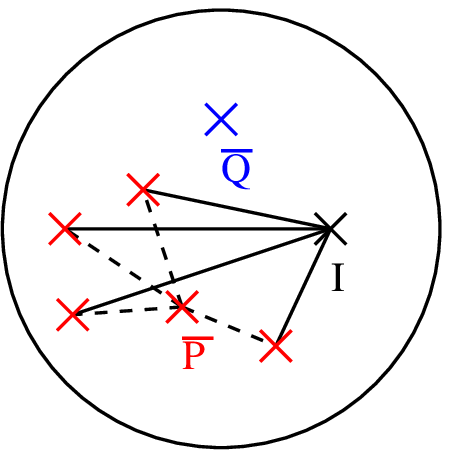}%
\caption{\label{fig:HBQ} Indirect heatbath updating generates $n_G$ proposals for plaquettes according to the canonical distribution whose projection is $\Pbar$.  Supposing $\Qbar \rightarrow \Pbar$ given fixed $\Gbar$ requires the updated link to take the value $U_\mrm{new} = \Pbar\, \Gbar^\dgr$.}
\end{figure}

The first step in developing a local heatbath (HB) algorithm is to break the relationship between the links $U_\mrm{new}$ and $U_\mrm{old}$ so that updates are generated with the canonical distribution $\rmd U_\mrm{new} \exp (- \beta S_\mrm{new})$ without regard to the current link value~\cite{Montvay:1994cy}.  As the direct generalization of the CKP heatbath algorithm is intractable, we propose an indirect scheme for generating the updates $U_\mrm{new}$.  We start by generating $n_G$ proposals for plaquette elements $P_k$ with action drawn from a gamma distribution as detailed below---one may think of these as selecting new environment elements for a fixed link value of $U_\mrm{old}$.  After taking their projection $\Pbar$, we reverse the sense of what has changed, supposing the environment projection remains $\Gbar$ while the plaquette projection $\Qbar \rightarrow \Pbar$, so that the updated link takes the value $U_\mrm{new} = \Pbar\, \Gbar^\dgr$, as shown in Fig.~\ref{fig:HBQ}.  While it would be nice to get $\Pbar$ with the generation of a single random element, the action of the projected sum does not appear to be simply related to the sum of the actions for $N > 2$.

\begin{figure}
\includegraphics[]{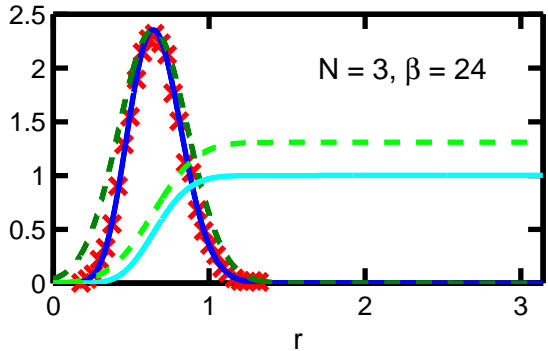}%
\caption{\label{fig:B} Normalized frequency and cumulative functions for the gamma (solid) and Gaussian (dashed) distributions as described in the text.  Also shown as $\times$ is the frequency distribution histogram of $10^4$ generated values. }
\end{figure}

The generation of the gamma distribution for the action $S_M (P_k)$ is accomplished using the rejection method~\cite{Press:1992}.  Retaining only the radial coordinate, the Haar measure~\cite{Montvay:1994cy} (accounting for our generator normalization) is $dP \propto dr\, r^{d-1} \exp [- r^2 N / 6 + \mathcal{O}(r^4)]$, and the local partition function $Z_P$ is written as \bea
\int dP\, e^{- \beta S_M (P)} &\propto& \int_0^{\norm{I'}_R} dr\, r^{d-1} e^{- (\beta / N + N / 6) r^2} \;, \\
 &\equiv& \int_0^{\norm{I'}_R} dr\, r^a e^{- b r^2} \;, \\
 &\propto& \gamma \left( \dfrac{a+1}{2}, b \norm{I'}_R^2 \right) \;, 
\eea where $\gamma (a,y) \equiv \int_0^y dr\, r^{a-1} e^{-r}$ is the incomplete gamma function.  The cumulative distribution is $F(r) = \gamma [(a+1)/2,b r^2] / Z_P$, and the unnormalized frequency distribution is $f(r) = r^a \exp (- b r^2) \propto dF / dr$.  The frequency distribution is peaked at $r_p = (a/2b)^{1/2}$ with a value $f_p = (a/2b)^{a/2} \exp (-a/2)$.  The distribution is approximated by first fitting a Gaussian to the range $r > r_p$ by minimizing $\int dr\, \lbrace \exp [-(r-r_p)^2 / 2 \sigma^2 ] - f / f_p \rbrace^2$ and then writing $g(r) = f_p \exp [-(r-r_p)^2 / s^2]$, where $s \equiv (3 \sigma^2)^{1/2}$ expands the Gaussian so that $g > f$.  Its integral yields the cumulative distribution \bea
G(r) &\equiv& \int_0^r dr'\, g(r') \;, \\
 &\propto& \mrm{erf}\! \left( \dfrac{r_p}{s} \right) - \mrm{erf}\! \left( \dfrac{r_p - r}{s} \right) \;,
\eea which one may invert analytically.  For a value $G_\mrm{draw}$ drawn uniformly from the range $[0, G(\norm{I'}_R)]$, one gets \beq
r_\mrm{draw} = r_p - s\; \mrm{erf}^{-1}\! \left[ \mrm{erf}\! \left( \dfrac{r_p}{s} \right)- G_\mrm{draw} \right] \;,
\eeq which is accepted with a probability of $f(r_\mrm{draw}) / g(r_\mrm{draw})$.  The normalized values of these distributions are shown in Fig.~\ref{fig:B} for $N=3$ and $\beta = 24$, along with the frequency distribution of $10^4$ draws.  Due to the close approximation $g \approx f$, there is a fairly high acceptance rate of $F_\mrm{max} / G_\mrm{max} \gtrsim 75\%$.

\begin{figure}
\includegraphics[]{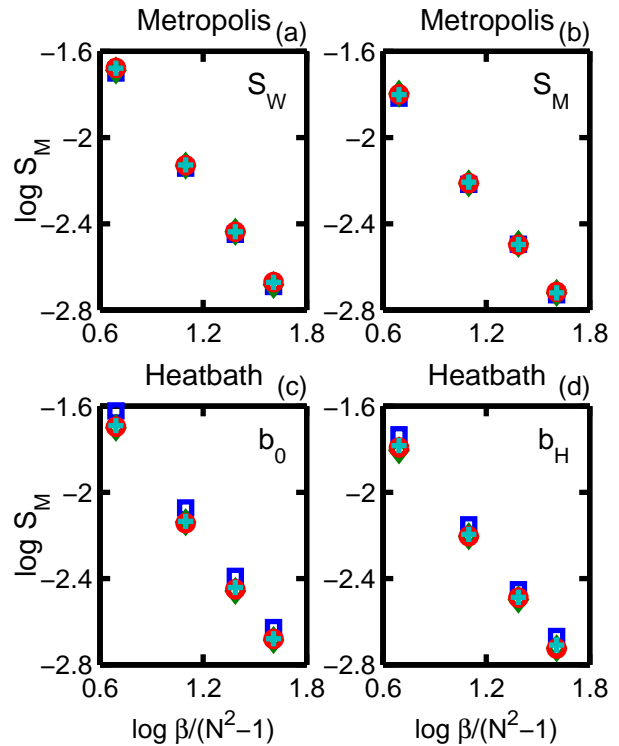}%
\caption{\label{fig:C} Comparison of the mean Manton action value $\Sbar_M$ for the Metropolis and indirect heatbath algorithms on a $L^D=4^3$ lattice as a function of the normalized coupling for $N \in \{2, 3, 4, 5\}$ plotted as $\{\Box, \Diamond, \bigcirc, +\}$. The mean action for MP using $S_W$ is higher than for MP using $S_M$, as is the value for HB using $b_0$ compared to using $b_H$.  The SU(2) values for HB are consistently higher than expected.}
\end{figure}

\section{\label{sec:distrib}Lattice action distribution}
To evaluate the performance of the indirect heatbath algorithm, we compare the expectation value of the plaquette action using the Manton definition $\Sbar_M \equiv \langle S_M(Q) \rangle_{\lbrace Q \rbrace}$ on a $D=3$, $L=4$ isotropic lattice measured over 100 configurations separated by 10 thermal sweeps following 100 thermalization sweeps from a warm start.  We consider the MP algorithm using either the Wilson or Manton action, and for the HB algorithm we compare the effect of neglecting the exponential factor of the Haar measure $b_0 = \beta / N$ with its inclusion $b_H = b_0 + N / 6$, as shown in Fig.~\ref{fig:C}.  Normalizing the coupling by the number of degrees of freedom, the $N$ dependence of the mean action disappears so that the values of $\Sbar_M (\beta / d)$ all follow the same linear relation on logarithmic axes---a similar plot obtains from the values presented by Teper~\cite{Teper:1999}.  The effect of the choice of action definition in the MP algorithm is most noticeable at low $\beta$ values, where $\Sbar_M (\mrm{MP}_W) > \Sbar_M (\mrm{MP}_M)$.  For the HB algorithm, we find that the mean action for SU(2) is consistently slightly higher than the corresponding values for $N>2$; we have not yet been able to root out the source of the discrepancy, but as the direct CKP algorithm applies here this issue is of no great concern.  Interestingly, we find agreement between using HB with $b_0$ and MP with $S_W$ and between using HB with $b_H$ and MP with $S_M$, which is understandable considering that both $b_H$ and $S_M$ include higher order corrections to $b_0$ and $S_W$ as elements get further from the identity.

\begin{figure}
\includegraphics[]{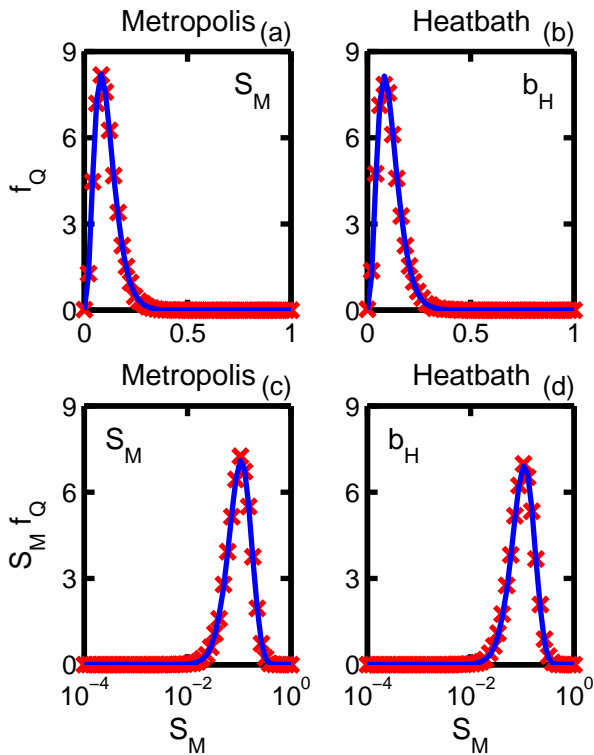}%
\caption{\label{fig:D} Frequency distribution of the Manton plaquette action for SU(3) at $\beta = 24$ plotted as histograms $\times$ with either linear or logarithmically spaced bins.  The regression function Eq.~\eref{eqn:fitfcn} is shown as the solid line.  The MP and HB algorithms return virtually identical action distributions.}
\end{figure}

We next compare the plaquette action distribution for SU(3) at $\beta = 24$ thermalized by MP with $S_M$ to that for HB with $b_H$, displayed in Fig.~\ref{fig:D} as histograms using either linear or logarithmically spaced bins.  The relation between the linear and logarithmic histograms may be expressed~\cite{Durrett:1994,Sivia:1996} as $f_Q dS_M = S_M f_Q d(\log S_M)$, and a least squares regression over the logarithmic bins of a function \beq \label{eqn:fitfcn}
f_Q \propto S_M^{-1+d/2} \exp \left( - \beta_\mrm{fit} S_M D / 2 \right)
\eeq yields an estimate of the coupling from the generated action distribution of $\beta_\mrm{fit}(\mrm{MP}_M) = 24.42$ and $\beta_\mrm{fit}(\mrm{HB}_H) = 24.24$.  (Technically one should perform a Bayesian regression to extract the parameters, but for our purposes here a simple fit is adequate.)  The fitting function also is shown in Fig.~\ref{fig:D}, where the agreement of the action distributions is apparent.

We use the fitting function Eq.~\eref{eqn:fitfcn} to extract the coupling from the generated distributions used in Fig.~\ref{fig:C} for $N \in \{2, 3, 4, 5\}$ and $\beta / d \in \{2, 3, 4, 5\}$, displayed in Table~\ref{tab:A}.  For consistency of comparison we use $S_M$ in Eq.~\eref{eqn:fitfcn} even when using $S_W$ or $b_0$ in the thermalization algorithm---these $\beta_\mrm{fit}$ are lower as $S_W < S_M$.  The $\beta_\mrm{fit}$ values for MP with $S_M$ and HB with $b_H$ are in excellent agreement with each other and with the coupling $\beta$ used in the updating algorithm, except for the case $N=2$ which seems to show a slight systematic discrepancy as mentioned above.  Certainly for the larger $N$, the Metropolis and indirect heatbath algorithms generate mutually compatible lattice action distributions when using the Manton action and accounting for the exponential factor of the Haar measure.

\begin{table}
\caption{\label{tab:A} Coupling values extracted using Eq.~\eref{eqn:fitfcn} from plaquette action distributions generated on a $L^D = 4^3$ lattice.}
\begin{ruledtabular}
\begin{tabular}{c|d{2}d{2}d{2}d{2}}
$N$ & \multicolumn{4}{c}{$\beta$} \\ \hline
2 &  6.00 &   9.00 &  12.00 &  15.00 \\
3 & 16.00 &  24.00 &  32.00 &  40.00 \\
4 & 30.00 &  45.00 &  60.00 &  75.00 \\
5 & 48.00 &  72.00 &  96.00 & 120.00 \\ \hline \hline
$N$ & \multicolumn{4}{c}{$\beta_\mrm{fit}$ for MP with $S_W$} \\ \hline
2 &  5.55 &   8.50 &  11.62 &  14.72 \\
3 & 14.48 &  22.52 &  30.61 &  39.12 \\
4 & 26.87 &  42.02 &  57.47 &  72.30 \\
5 & 42.96 &  67.30 &  91.56 & 116.10 \\ \hline \hline
$N$ & \multicolumn{4}{c}{$\beta_\mrm{fit}$ for HB with $b_0$} \\ \hline
2 &  5.11 &   7.99 &  11.07 &  13.85 \\
3 & 14.65 &  22.78 &  31.17 &  39.23 \\
4 & 27.38 &  42.60 &  58.24 &  73.19 \\
5 & 43.44 &  67.80 &  92.29 & 117.10 \\ \hline \hline
$N$ & \multicolumn{4}{c}{$\beta_\mrm{fit}$ for MP with $S_M$} \\ \hline
2 &  6.13 &   9.30 &  12.14 &  15.27 \\
3 & 16.08 &  24.42 &  32.26 &  40.77 \\
4 & 30.28 &  45.81 &  60.93 &  75.75 \\
5 & 48.62 &  72.78 &  97.22 & 121.67 \\ \hline \hline
$N$ & \multicolumn{4}{c}{$\beta_\mrm{fit}$ for HB with $b_H$} \\ \hline
2 &  5.70 &   8.65 &  11.70 &  14.52 \\
3 & 16.28 &  24.24 &  32.57 &  40.86 \\
4 & 30.13 &  45.36 &  60.66 &  76.79 \\
5 & 47.90 &  72.07 &  96.53 & 120.42 
\end{tabular}
\end{ruledtabular}
\end{table}

\section{\label{sec:rates}Running times and decorrelation rates}
\begin{figure*}
\includegraphics[]{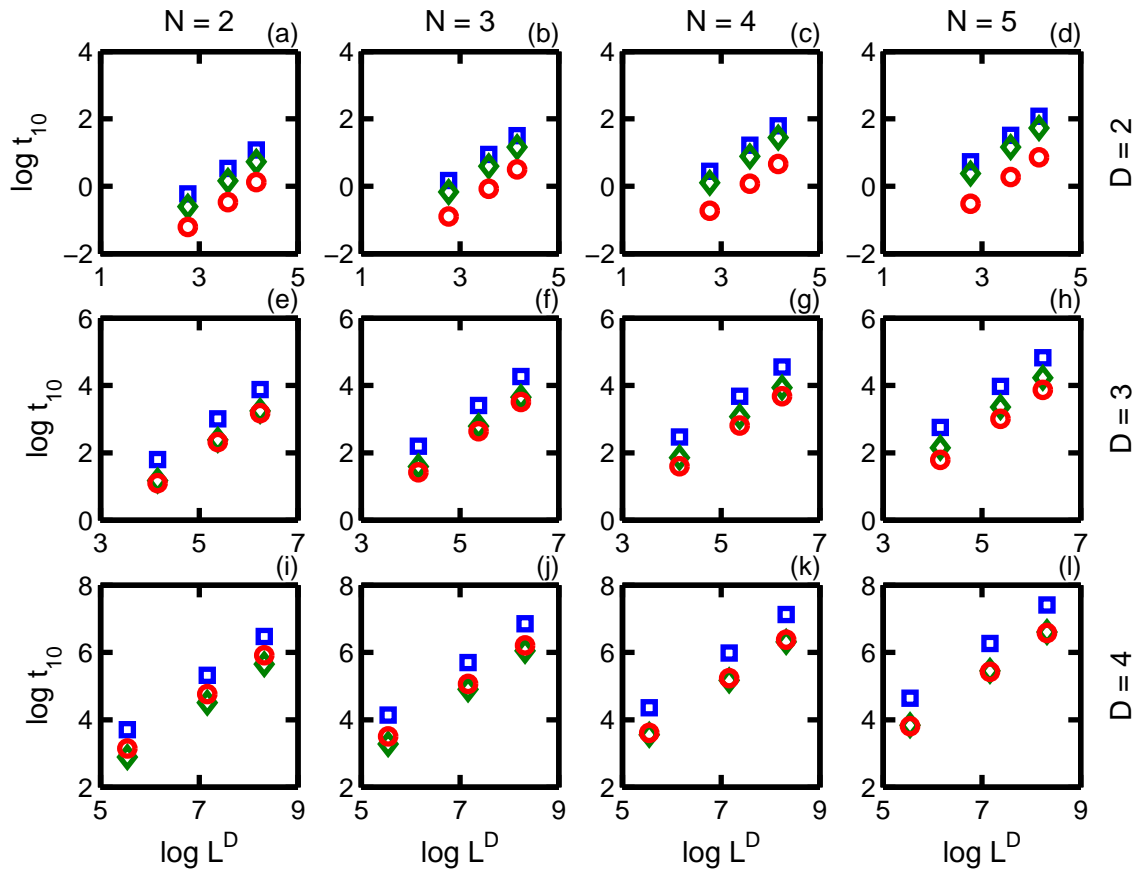}%
\caption{\label{fig:E} Running times $t_{10}$ for 10 thermal updating sweeps as a function of lattice volume $L^D$ for MP with $S_M$ shown as $\Box$, for MP with $S_W$ as $\Diamond$, and for HB with $b_H$ as $\bigcirc$. }
\end{figure*}

To compare the efficiency of the algorithms, we measure the running time $t_{10}$ of 10 thermalization sweeps for lattices with $L \in \{4, 6, 8\}$ and $D \in \{2, 3, 4\}$, as shown in Fig.~\ref{fig:E}.  No optimizations have been made beyond the obvious, and the evaluation is on a serial processor.  The expected linear dependence on lattice volume $t_{10} \propto L^D$ is observed.  The ratio of the running time for the MP algorithm using the Manton action to that for the Wilson action increases along with the lattice dimension $D$.  The running time for HB (using $b_H$) is always less than that for MP using $S_M$ and is usually less than that for MP using $S_W$.  The indirect heatbath algorithm allows for the use of the Manton action with a computational efficiency exceeding that of the corresponding Metropolis algorithm.

To measure the correlation between two configurations $\{U(x)\}$ and $\{U'(x)\}$ separated by a given number of update sweeps, we consider both the link correlation $C(U,U') \equiv 1 - \Vert U^\dgr U' \Vert_R / \Vert I' \Vert_R$ and the plaquette correlation $C(Q,Q') \equiv 1 - \Vert Q^\dgr Q' \Vert_R / \Vert I' \Vert_R$ evaluated for a particular site and averaged over the lattice directions and volume.  This measure of correlation equals 1 when $U = U'$ and descends to some residual value determined by the inverse statistical temperature, as for high $\beta$ none of the elements may stray far from the identity---early on we reproduced values similar to those of Ref.~\cite{Creutz:1987xi} using that definition of correlation.  

\begin{figure}
\includegraphics[]{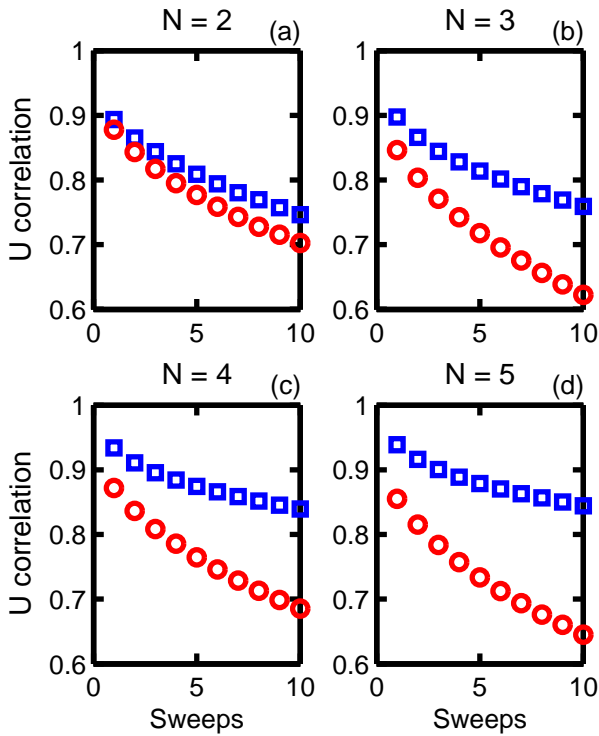}%
\caption{\label{fig:F} Link correlation $C(U,U')$ as a function of the number of updating sweeps at $\beta / d = 3$ on a $L^D = 4^3$ lattice for the MP $\Box$ and the HB $\bigcirc$ algorithms. }
\end{figure}

\begin{figure}
\includegraphics[]{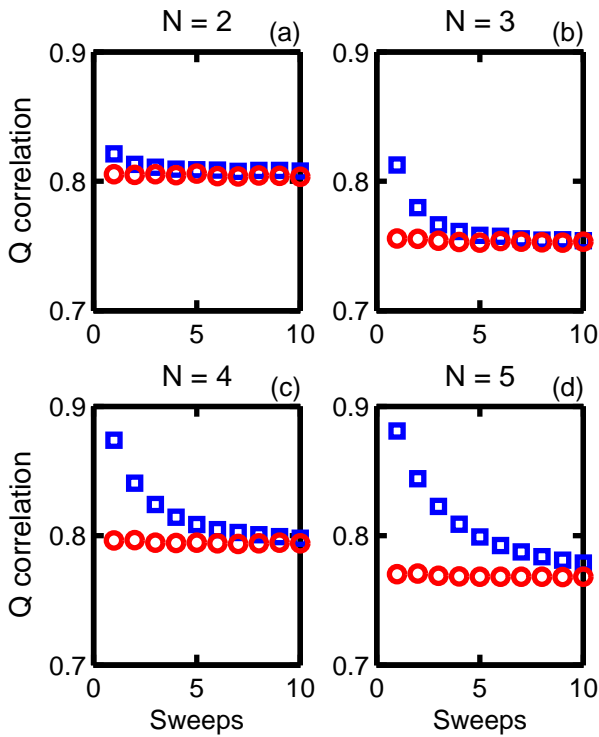}%
\caption{\label{fig:G} Plaquette correlation $C(Q,Q')$ as a function of the number of updating sweeps at $\beta / d = 3$ on a $L^D = 4^3$ lattice for the MP $\Box$ and the HB $\bigcirc$ algorithms. }
\end{figure}

We evaluate the link and plaquette correlations on a $L^D = 4^3$ lattice with $\beta / d = 3$, comparing MP with $S_M$ to HB with $b_H$ averaged over 100 measurements.  Only for $N=5$ was a slight volume dependence seen for MP compared to a short run with $L=6$.  The link correlation $C(U,U')$ decays much faster for the HB algorithm, as shown in Fig.~\ref{fig:F}, where the rate of decorrelation displays a dependence on the gauge group $N$.  As links are not gauge invariant objects, we also measure the plaquette correlation $C(Q,Q')$, shown in Fig.~\ref{fig:G}.  Here we see that the HB algorithm has achieved nearly the minimal correlation after only a single sweep, in sharp contrast to the MP algorithm which requires a greater number of sweeps to reach minimal correlation as $N$ increases.  Having broken the relationship between the links $U_\mrm{new}$ and $U_\mrm{old}$ in the thermal updating, the indirect heatbath algorithm outperforms the Metropolis algorithm at decorrelating the lattice gauge field configurations.

\section{\label{sec:plaqrep}Link {\it vs} plaquette simulation}
As an interesting conjoinder to our investigation of link updating by the heatbath algorithm, we would like to present some thoughts on an alternative approach to quantum gauge field simulations on a Euclidean lattice.  An efficient means of generating random elements $Q$ with known action $S_M(Q)$ suggests the possibility of running a lattice simulation with plaquettes rather than links as the dynamic variables.  As there is a $1:1$ geometrical correspondence between the number of links and the number of plaquettes, a given lattice configuration $\{U\}$ may equally well be represented as a configuration $\{Q\}$.  Furthermore, the representation in $\{Q\}$ is gauge invariant, corresponding to an entire class of $\{U\}$ related by gauge transformations.  Essentially, the plaquette representation ``integrates out'' the gauge degrees of freedom to leave one with invariant SU($N$) elements.  How the lattice Bianchi identities should be satisfied needs to be investigated~\cite{borisenko:2009399}.

The obvious difficulty with such an approach is the construction of operators---how should they appear in the plaquette representation?  We conjecture that the building blocks of plaquette elements should be added together to form closed loops or bags of plaquettes, much as links are multiplied in their representation, so that all boundaries are closed (in the pure gauge theory).  That construction is not unlike the depiction of bound currents in the macroscopic treatment of electromagnetism, a topic which may be familiar to some readers~\cite{CambridgeJournals:6902496}.  For the sectors of $J^{PC}$ with $C=+$ for $N \geq 2$, the usual operator $\Re \Tr Q_\mathcal{C}$ built from the trace of a product of links $Q_\mathcal{C} \equiv  \Pi_\mathcal{C} U(\mathcal{C})$ over a closed path $\mathcal{C}$ is obviously related to the Manton action for the element $Q_\mathcal{C}$, but for $C=-$ a similar construction is not yet known to us.  Investigation of this approach is beyond the scope of the present work but remains a topic of interest.

\section{\label{sec:concl}Conclusions}
The availability of a general heatbath algorithm for SU($N$) lattice gauge theory has long been desired.  While the Metropolis algorithm certainly is effective at thermalization, its implementation requires an overcommitment of resources to achieve a significant level of decorrelation.  The advantage of the indirect heatbath algorithm is that a fewer number of trials are proposed for each link in a sweep, with consequent savings in time, energy, and ultimately money, as each flop exacts a price however small from the investigator.  While we have not compared the indirect heatbath algorithm directly to that by Cabibbo and Marinari~\cite{Cabibbo1982387}, we expect to see a similar savings due to treating each element as a unit rather than decomposing it according to a covering set of subgroups.

From the identification of the lattice action as the squared distance from the identity of the invariant plaquette element we have constructed a general heatbath algorithm for SU($N$) pure gauge theory.  The algorithm consists of proposing a number of new plaquette elements, whose mean yields the updated link after dividing out the projection of the surrounding staple elements.  The internal energy distribution obtained by the heatbath algorithm agrees with that produced by Metropolis updating, and the link and plaquette decorrelation rates compare favorably, as does the execution speed.  The indirect heatbath algorithm incorporates the Manton action, to which the Wilson action is a leading order approximation, accounting for the metric within the special unitary group rather than the general linear group.  By breaking the relationship between the link to be updated and its replacement, the indirect heatbath algorithm achieves the maximal decorrelation of gauge invariant objects with a minimal amount of effort.


\vspace{-0.3cm}
\begin{acknowledgments}
The author appreciates occasional conversations with Mike Teper on the use of lattice gauge theory, particularly for the suggestion to check the histograms, and with Philippe de Forcrand on various details of thermal updating.
\end{acknowledgments}




\begin{thebibliography}{21}%
\makeatletter
\providecommand \@ifxundefined [1]{%
 \@ifx{#1\undefined}
}%
\providecommand \@ifnum [1]{%
 \ifnum #1\expandafter \@firstoftwo
 \else \expandafter \@secondoftwo
 \fi
}%
\providecommand \@ifx [1]{%
 \ifx #1\expandafter \@firstoftwo
 \else \expandafter \@secondoftwo
 \fi
}%
\providecommand \natexlab [1]{#1}%
\providecommand \enquote  [1]{``#1''}%
\providecommand \bibnamefont  [1]{#1}%
\providecommand \bibfnamefont [1]{#1}%
\providecommand \citenamefont [1]{#1}%
\providecommand \href@noop [0]{\@secondoftwo}%
\providecommand \href [0]{\begingroup \@sanitize@url \@href}%
\providecommand \@href[1]{\@@startlink{#1}\@@href}%
\providecommand \@@href[1]{\endgroup#1\@@endlink}%
\providecommand \@sanitize@url [0]{\catcode `\\12\catcode `\$12\catcode
  `\&12\catcode `\#12\catcode `\^12\catcode `\_12\catcode `\%12\relax}%
\providecommand \@@startlink[1]{}%
\providecommand \@@endlink[0]{}%
\providecommand \url  [0]{\begingroup\@sanitize@url \@url }%
\providecommand \@url [1]{\endgroup\@href {#1}{\urlprefix }}%
\providecommand \urlprefix  [0]{URL }%
\providecommand \Eprint [0]{\href }%
\providecommand \doibase [0]{http://dx.doi.org/}%
\providecommand \selectlanguage [0]{\@gobble}%
\providecommand \bibinfo  [0]{\@secondoftwo}%
\providecommand \bibfield  [0]{\@secondoftwo}%
\providecommand \translation [1]{[#1]}%
\providecommand \BibitemOpen [0]{}%
\providecommand \bibitemStop [0]{}%
\providecommand \bibitemNoStop [0]{.\EOS\space}%
\providecommand \EOS [0]{\spacefactor3000\relax}%
\providecommand \BibitemShut  [1]{\csname bibitem#1\endcsname}%
\let\auto@bib@innerbib\@empty
\bibitem [{\citenamefont {Manton}(1980)}]{Manton:1980328}%
  \BibitemOpen
  \bibfield  {author} {\bibinfo {author} {\bibfnamefont {N.}~\bibnamefont
  {Manton}},\ }\href {\doibase DOI: 10.1016/0370-2693(80)90778-9} {\bibfield
  {journal} {\bibinfo  {journal} {Physics Letters B}\ }\textbf {\bibinfo
  {volume} {96}},\ \bibinfo {pages} {328 } (\bibinfo {year}
  {1980})}\BibitemShut {NoStop}%
\bibitem [{\citenamefont {Moakher}(2002)}]{moakher:02895}%
  \BibitemOpen
  \bibfield  {author} {\bibinfo {author} {\bibfnamefont {M.}~\bibnamefont
  {Moakher}},\ }\href {\doibase http://dx.doi.org/10.1137/S0895479801383877}
  {\bibfield  {journal} {\bibinfo  {journal} {SIAM J. Matrix Anal. Appl.}\
  }\textbf {\bibinfo {volume} {24}},\ \bibinfo {pages} {1} (\bibinfo {year}
  {2002})}\BibitemShut {NoStop}%
\bibitem [{\citenamefont {Bazavov}\ and\ \citenamefont
  {Berg}(2005)}]{bazavov:114506}%
  \BibitemOpen
  \bibfield  {author} {\bibinfo {author} {\bibfnamefont {A.}~\bibnamefont
  {Bazavov}}\ and\ \bibinfo {author} {\bibfnamefont {B.~A.}\ \bibnamefont
  {Berg}},\ }\href {\doibase 10.1103/PhysRevD.71.114506} {\bibfield  {journal}
  {\bibinfo  {journal} {Phys. Rev. D}\ }\textbf {\bibinfo {volume} {71}},\
  \bibinfo {pages} {114506} (\bibinfo {year} {2005})}\BibitemShut {NoStop}%
\bibitem [{\citenamefont {Bazavov}\ \emph {et~al.}(2005)\citenamefont
  {Bazavov}, \citenamefont {Berg},\ and\ \citenamefont
  {Heller}}]{bazavov:117501}%
  \BibitemOpen
  \bibfield  {author} {\bibinfo {author} {\bibfnamefont {A.}~\bibnamefont
  {Bazavov}}, \bibinfo {author} {\bibfnamefont {B.~A.}\ \bibnamefont {Berg}}, \
  and\ \bibinfo {author} {\bibfnamefont {U.~M.}\ \bibnamefont {Heller}},\
  }\href {\doibase 10.1103/PhysRevD.72.117501} {\bibfield  {journal} {\bibinfo
  {journal} {Phys. Rev. D}\ }\textbf {\bibinfo {volume} {72}},\ \bibinfo
  {pages} {117501} (\bibinfo {year} {2005})}\BibitemShut {NoStop}%
\bibitem [{\citenamefont {Creutz}(1980)}]{Creutz:1980mc}%
  \BibitemOpen
  \bibfield  {author} {\bibinfo {author} {\bibfnamefont {M.}~\bibnamefont
  {Creutz}},\ }\href {\doibase 10.1103/PhysRevD.21.2308} {\bibfield  {journal}
  {\bibinfo  {journal} {Phys. Rev. D}\ }\textbf {\bibinfo {volume} {21}},\
  \bibinfo {pages} {2308} (\bibinfo {year} {1980})}\BibitemShut {NoStop}%
\bibitem [{\citenamefont {Kennedy}\ and\ \citenamefont
  {Pendleton}(1985)}]{Kennedy:1985nu}%
  \BibitemOpen
  \bibfield  {author} {\bibinfo {author} {\bibfnamefont {A.~D.}\ \bibnamefont
  {Kennedy}}\ and\ \bibinfo {author} {\bibfnamefont {B.~J.}\ \bibnamefont
  {Pendleton}},\ }\href@noop {} {\bibfield  {journal} {\bibinfo  {journal}
  {Physics Letters B}\ }\textbf {\bibinfo {volume} {156}},\ \bibinfo {pages}
  {393} (\bibinfo {year} {1985})}\BibitemShut {NoStop}%
\bibitem [{\citenamefont {Cabibbo}\ and\ \citenamefont
  {Marinari}(1982)}]{Cabibbo1982387}%
  \BibitemOpen
  \bibfield  {author} {\bibinfo {author} {\bibfnamefont {N.}~\bibnamefont
  {Cabibbo}}\ and\ \bibinfo {author} {\bibfnamefont {E.}~\bibnamefont
  {Marinari}},\ }\href {\doibase DOI: 10.1016/0370-2693(82)90696-7} {\bibfield
  {journal} {\bibinfo  {journal} {Physics Letters B}\ }\textbf {\bibinfo
  {volume} {119}},\ \bibinfo {pages} {387 } (\bibinfo {year}
  {1982})}\BibitemShut {NoStop}%
\bibitem [{\citenamefont {Teper}(1999)}]{Teper:1999}%
  \BibitemOpen
  \bibfield  {author} {\bibinfo {author} {\bibfnamefont {M.}~\bibnamefont
  {Teper}},\ }\href@noop {} {\bibfield  {journal} {\bibinfo  {journal}
  {Physical Review D}\ }\textbf {\bibinfo {volume} {59}} (\bibinfo {year}
  {1999})},\ \Eprint {http://arxiv.org/abs/hep-lat/9804008} {hep-lat/9804008}
  \BibitemShut {NoStop}%
\bibitem [{\citenamefont {Metropolis}\ \emph {et~al.}(1953)\citenamefont
  {Metropolis}, \citenamefont {Rosenbluth}, \citenamefont {Rosenbluth},
  \citenamefont {Teller},\ and\ \citenamefont {Teller}}]{metropolis:1087}%
  \BibitemOpen
  \bibfield  {author} {\bibinfo {author} {\bibfnamefont {N.}~\bibnamefont
  {Metropolis}}, \bibinfo {author} {\bibfnamefont {A.~W.}\ \bibnamefont
  {Rosenbluth}}, \bibinfo {author} {\bibfnamefont {M.~N.}\ \bibnamefont
  {Rosenbluth}}, \bibinfo {author} {\bibfnamefont {A.~H.}\ \bibnamefont
  {Teller}}, \ and\ \bibinfo {author} {\bibfnamefont {E.}~\bibnamefont
  {Teller}},\ }\href {\doibase 10.1063/1.1699114} {\bibfield  {journal}
  {\bibinfo  {journal} {The Journal of Chemical Physics}\ }\textbf {\bibinfo
  {volume} {21}},\ \bibinfo {pages} {1087} (\bibinfo {year}
  {1953})}\BibitemShut {NoStop}%
\bibitem [{\citenamefont {Montvay}\ and\ \citenamefont
  {Munster}(1994)}]{Montvay:1994cy}%
  \BibitemOpen
  \bibfield  {author} {\bibinfo {author} {\bibfnamefont {I.}~\bibnamefont
  {Montvay}}\ and\ \bibinfo {author} {\bibfnamefont {G.}~\bibnamefont
  {Munster}},\ }\href@noop {} {\emph {\bibinfo {title} {Quantum Fields on a
  Lattice}}}\ (\bibinfo  {publisher} {CUP},\ \bibinfo {address} {Cambridge,
  England},\ \bibinfo {year} {1994})\ \bibinfo {note} {{C}ambridge monographs
  on mathematical physics}\BibitemShut {NoStop}%
\bibitem [{\citenamefont {Morningstar}\ and\ \citenamefont
  {Peardon}(2004)}]{morning:04695}%
  \BibitemOpen
  \bibfield  {author} {\bibinfo {author} {\bibfnamefont {C.}~\bibnamefont
  {Morningstar}}\ and\ \bibinfo {author} {\bibfnamefont {M.}~\bibnamefont
  {Peardon}},\ }\href {\doibase 10.1103/PhysRevD.69.054501} {\bibfield
  {journal} {\bibinfo  {journal} {Phys. Rev. D}\ }\textbf {\bibinfo {volume}
  {69}},\ \bibinfo {pages} {054501} (\bibinfo {year} {2004})}\BibitemShut
  {NoStop}%
\bibitem [{\citenamefont {Liang}\ \emph {et~al.}(1993)\citenamefont {Liang},
  \citenamefont {Liu}, \citenamefont {Li}, \citenamefont {Dong},\ and\
  \citenamefont {Ishikawa}}]{Liang:1992cz}%
  \BibitemOpen
  \bibfield  {author} {\bibinfo {author} {\bibfnamefont {Y.}~\bibnamefont
  {Liang}}, \bibinfo {author} {\bibfnamefont {K.~F.}\ \bibnamefont {Liu}},
  \bibinfo {author} {\bibfnamefont {B.~A.}\ \bibnamefont {Li}}, \bibinfo
  {author} {\bibfnamefont {S.~J.}\ \bibnamefont {Dong}}, \ and\ \bibinfo
  {author} {\bibfnamefont {K.}~\bibnamefont {Ishikawa}},\ }\href@noop {}
  {\bibfield  {journal} {\bibinfo  {journal} {Phys. Lett.}\ }\textbf {\bibinfo
  {volume} {B307}},\ \bibinfo {pages} {375} (\bibinfo {year} {1993})},\ \Eprint
  {http://arxiv.org/abs/hep-lat/9304011} {hep-lat/9304011} \BibitemShut
  {NoStop}%
\bibitem [{\citenamefont {Creutz}(1987)}]{Creutz:1987xi}%
  \BibitemOpen
  \bibfield  {author} {\bibinfo {author} {\bibfnamefont {M.}~\bibnamefont
  {Creutz}},\ }\href {\doibase 10.1103/PhysRevD.36.515} {\bibfield  {journal}
  {\bibinfo  {journal} {Phys. Rev. D}\ }\textbf {\bibinfo {volume} {36}},\
  \bibinfo {pages} {515} (\bibinfo {year} {1987})}\BibitemShut {NoStop}%
\bibitem [{\citenamefont {Adler}(1988)}]{Adler:37458}%
  \BibitemOpen
  \bibfield  {author} {\bibinfo {author} {\bibfnamefont {S.~L.}\ \bibnamefont
  {Adler}},\ }\href {\doibase 10.1103/PhysRevD.37.458} {\bibfield  {journal}
  {\bibinfo  {journal} {Phys. Rev. D}\ }\textbf {\bibinfo {volume} {37}},\
  \bibinfo {pages} {458} (\bibinfo {year} {1988})}\BibitemShut {NoStop}%
\bibitem [{\citenamefont {Forcrand}\ and\ \citenamefont
  {Jahn}(2005)}]{forcrand:28504}%
  \BibitemOpen
  \bibfield  {author} {\bibinfo {author} {\bibfnamefont {P.}~\bibnamefont
  {Forcrand}}\ and\ \bibinfo {author} {\bibfnamefont {O.}~\bibnamefont
  {Jahn}},\ }in\ \href {\doibase 10.1007/3-540-28504-0_6} {\emph {\bibinfo
  {booktitle} {QCD and Numerical Analysis III}}},\ \bibinfo {series} {Lecture
  Notes in Computational Science and Engineering}, Vol.~\bibinfo {volume}
  {47},\ \bibinfo {editor} {edited by\ \bibinfo {editor} {\bibfnamefont
  {A.}~\bibnamefont {Bori~i}}, \bibinfo {editor} {\bibfnamefont
  {A.}~\bibnamefont {Frommer}}, \bibinfo {editor} {\bibfnamefont
  {B.}~\bibnamefont {Joó}}, \bibinfo {editor} {\bibfnamefont {A.}~\bibnamefont
  {Kennedy}}, \ and\ \bibinfo {editor} {\bibfnamefont {B.}~\bibnamefont
  {Pendleton}}}\ (\bibinfo  {publisher} {Springer Berlin Heidelberg},\ \bibinfo
  {year} {2005})\ pp.\ \bibinfo {pages} {67--73}\BibitemShut {NoStop}%
\bibitem [{\citenamefont {{Weigert}}(1997)}]{weigert:8739W}%
  \BibitemOpen
  \bibfield  {author} {\bibinfo {author} {\bibfnamefont {S.}~\bibnamefont
  {{Weigert}}},\ }\href {\doibase 10.1088/0305-4470/30/24/032} {\bibfield
  {journal} {\bibinfo  {journal} {J. Phys. A: Math. Gen.}\ }\textbf {\bibinfo
  {volume} {30}},\ \bibinfo {pages} {8739} (\bibinfo {year} {1997})},\ \Eprint
  {http://arxiv.org/abs/arXiv:quant-ph/9710024} {arXiv:quant-ph/9710024}
  \BibitemShut {NoStop}%
\bibitem [{\citenamefont {Press}\ \emph {et~al.}(1992)\citenamefont {Press},
  \citenamefont {Teukolsky}, \citenamefont {Vetterling},\ and\ \citenamefont
  {Flannery}}]{Press:1992}%
  \BibitemOpen
  \bibfield  {author} {\bibinfo {author} {\bibfnamefont {W.}~\bibnamefont
  {Press}}, \bibinfo {author} {\bibfnamefont {S.}~\bibnamefont {Teukolsky}},
  \bibinfo {author} {\bibfnamefont {W.}~\bibnamefont {Vetterling}}, \ and\
  \bibinfo {author} {\bibfnamefont {B.}~\bibnamefont {Flannery}},\ }\href@noop
  {} {\emph {\bibinfo {title} {Numerical Recipes in C}}},\ \bibinfo {edition}
  {2nd}\ ed.\ (\bibinfo  {publisher} {Cambridge University Press},\ \bibinfo
  {address} {Cambridge, England},\ \bibinfo {year} {1992})\BibitemShut
  {NoStop}%
\bibitem [{\citenamefont {{Durrett}}(1994)}]{Durrett:1994}%
  \BibitemOpen
  \bibfield  {author} {\bibinfo {author} {\bibfnamefont {R.}~\bibnamefont
  {{Durrett}}},\ }\href@noop {} {\emph {\bibinfo {title} {The Essentials of
  Probability}}}\ (\bibinfo  {publisher} {Duxbury Press, A Division of
  Wadsworth, Inc.},\ \bibinfo {address} {Belmont, California, USA},\ \bibinfo
  {year} {1994})\BibitemShut {NoStop}%
\bibitem [{\citenamefont {Sivia}(1996)}]{Sivia:1996}%
  \BibitemOpen
  \bibfield  {author} {\bibinfo {author} {\bibfnamefont {D.~S.}\ \bibnamefont
  {Sivia}},\ }\href@noop {} {\emph {\bibinfo {title} {Data Analysis: a Bayesian
  Primer}}}\ (\bibinfo  {publisher} {OUP},\ \bibinfo {address} {Oxford,
  England},\ \bibinfo {year} {1996})\BibitemShut {NoStop}%
\bibitem [{\citenamefont {Borisenko}\ \emph {et~al.}(2009)\citenamefont
  {Borisenko}, \citenamefont {Voloshin},\ and\ \citenamefont
  {Faber}}]{borisenko:2009399}%
  \BibitemOpen
  \bibfield  {author} {\bibinfo {author} {\bibfnamefont {O.}~\bibnamefont
  {Borisenko}}, \bibinfo {author} {\bibfnamefont {S.}~\bibnamefont {Voloshin}},
  \ and\ \bibinfo {author} {\bibfnamefont {M.}~\bibnamefont {Faber}},\ }\href
  {\doibase DOI: 10.1016/j.nuclphysb.2009.02.008} {\bibfield  {journal}
  {\bibinfo  {journal} {Nuclear Physics B}\ }\textbf {\bibinfo {volume}
  {816}},\ \bibinfo {pages} {399 } (\bibinfo {year} {2009})}\BibitemShut
  {NoStop}%
\bibitem [{\citenamefont {Johnson}(2009)}]{CambridgeJournals:6902496}%
  \BibitemOpen
  \bibfield  {author} {\bibinfo {author} {\bibfnamefont {R.~W.}\ \bibnamefont
  {Johnson}},\ }\href {\doibase 10.1017/S002237780999050X} {\bibfield
  {journal} {\bibinfo  {journal} {Journal of Plasma Physics}\ }\textbf
  {\bibinfo {volume} {First View}},\ \bibinfo {pages} {1} (\bibinfo {year}
  {2009})}\BibitemShut {NoStop}%
\end{thebibliography}
%

\end{document}